\documentstyle[prl,aps,epsf]{revtex}

\begin{document} 
\draft
\title{Berry phase and persistent current in disordered mesoscopic rings}
\author{Shiro Kawabata\thanks {E-mail: shiro@etl.go.jp}}
\address{
Physical Science Division, Electrotechnical Laboratory, 1-1-4 Umezono, Tsukuba, 
Ibaraki 305-8568, Japan
}
%
\date{\today}
\maketitle
\begin{abstract} 

A novel quantum interference effect in disordered quasi-one-dimensional rings in the inhomogeneous magnetic field is reported.
We calculate the canonical disorder averaged persistent current using the diagrammatic perturbation theory.
It is shown that within the adiabatic regime the average current oscillates as a function of the geometric flux which is related  to the Berry phase and the period becomes half the value of the case of a single one-dimensional ring. 
We also discuss the magnetic dephasing effect on the averaged current.

\end{abstract}
%
%
%
%
%
%
%
%
%
%
%
%
%
Since the discovery of the Berry phase,~\cite{rf:Berry} there has been much interest in the study of topological effects in the fields of quantum mechanics and condensed matter physics.~\cite{rf:BerryRev}
The typical example to illustrate the Berry phase is the Aharonov-Bohm (AB) effect~\cite{rf:AB} in the mesoscopic ring, where the relative phase would accumulate on the wave function of a charged particle due to the presence of a electromagnetic gauge potential.
Similarly, when a quantum spin follows adiabatically a magnetic field that rotates slowly in time, the spin wave function acquires an additional geometric phase (Berry phase) besides the usual electromagnetic phase in the static magnetic fields.

Loss $et$ $al$.~\cite{rf:LossPC} studied the persistent current in an $ideal$ one-dimensional ring in the presence of a static inhomogeneous magnetic field by examining the coupling between spin and orbital motion through the Zeeman interaction.
Using the imaginary time path integral method, they showed that the spin wave function accumulates the Berry phase when the spin of an electron traversing an AB ring adiabatically follows a textured inhomogeneous magnetic filed with a tilt angle and this phase leads to persistent equilibrium charge current. 

In this paper we investigate the persistent charge current of the  quasi-one-dimensional $disordered$ rings~\cite{rf:PCexp} (see the inset in Fig. 2) and compare it with the results of Loss $et$ $al$.~\cite{rf:LossPC}
We compute the disorder averaged persistent current with the help of the diagrammatic technique and find that the average current is given by the sum of the term which oscillates as a function of the Berry phase and the term which does not depend on the the Berry phase.
We show that the latter term is negligibly small compared with the former term in the adiabatic regime, so that we will be able to clearly observe the Berry phase dependent persistent current experimentally.

We begin by considering a quasi-one-dimensional ring of circumference $L_x=2 \pi r$ and volume $V=L_xL_yL_z$.
The ring is embedded in an static inhomogeneous  magnetic field ${\bf B}$.
For a spin-$1/2$ electrons of mass $m$ and charge $e$, the system may be described by the Hamiltonian
\begin{eqnarray}
{\cal H}
      =
      \frac{1}{2m} \left[  {\bf p} - \frac{e}{c} {\bf A}^{em}({\bf r}) \right]^2 
     +u({\bf r} ) 
	 -\frac{1}{2} g \mu_B {\bf B} ({\bf r} ) 
	 \mbox{\boldmath $\cdot$}
	 \mbox{\boldmath $\sigma$}
	 ,
  \label{eqn:e1}
\end{eqnarray}
where ${\bf p}$, ${\bf r}$, $g$, $\mu_B$, and $\hbar \mbox{\boldmath $\sigma$}/2$ are the momentum, position, $g$ factor, Bohr magneton, and spin, respectively.
The operator $u({\bf r})$ represents the spin-independent random impurity potential and, ${\bf A}^{em}$ is the electromagnetic gauge potential, with ${\bf B}=\nabla \times {\bf A}^{em}$ relating it to the magnetic fields.
In the following we specialize in the case of inhomogeneous magnetic fields with constant magnitude $B$, and we have parametrized ${\bf B}$ in terms of the spherical polar angles $\chi$ and $\eta$ so that it has Cartesian components $B\left(  \sin \chi({\bf r}) \cos \eta({\bf r}),  \sin \chi({\bf r}) \sin \eta({\bf r}), \cos \chi({\bf r}) \right)$, with the angles $\chi$ and $\eta$ being smooth functions of position.

Using the Green's function, the thermal average of the persistent current is given by
\begin{equation}
 I=\int \frac{d\varepsilon}{2 \pi i} \sum_{{\bf k}} \sum_{\alpha=\pm 1}
                 I_x f(\varepsilon) \left[ G^{A}_{\alpha, \alpha} ({\bf k},{\bf k},\varepsilon)
				                           -
										   G^{R}_{\alpha, \alpha} ({\bf k},{\bf k},\varepsilon)
								\right]
	,
  \label{eqn:e2}
\end{equation}
where $\alpha=\pm 1$ is the spin index and  $f(\varepsilon)$ is the Fermi-Dirac distribution function.
The current vertex is given by $I_x=-e \hbar k_x/mL$, where $k_x$ is the $x$ component of the wave number vector.
Moreover $G^{R(A)}$ is the retarded (advanced) Green's function and it denotes the exact Green's function before impurity averaging.
To determine the canonical disorder-averaged persistent current, we follow the procedure described in Refs. [6$\sim$8] 
The chemical potential  is set to $\mu=\mu_0 + \delta \mu$ and we expand Eq.~(\ref{eqn:e2}) to first order in $\delta \mu$.
Then we obtain the averaged  persistent current at a fixed particle number as follows:
\begin{equation}
      \left< I(\Phi^{em}) \right>
	  \approx 
	  - \frac{\Delta}{2} 
         \frac{ \partial}{\partial \Phi^{em}} 
		 \left<
	     \left\{  
		 \sum_{\alpha=\pm 1} \delta N_{\alpha}(\mu_0,\Phi^{em})
		 \right\}^2
		 \right>
		 ,
  \label{eqn:e3}
\end{equation}
where $\Phi^{em}$ is the electromagnetic flux and $\Delta $ is the mean level spacing.
The fluctuation of the particle number $\left<  (\sum\nolimits_\alpha \delta N_{\alpha})^2 \right>$ is related to the two-point correlator of the density of states $\rho_\alpha$,~\cite{rf:AltshulerShklovsky}
\begin{equation}
		 \left<
	     \left\{  
		 \sum_{\alpha=\pm 1} \delta N_{\alpha}(\mu_0,\Phi^{em})
		 \right\}^2
		 \right>
	  = 
	  V^2
	  \int_{-\infty}^{\infty} d \varepsilon_1 
	  \int_{-\infty}^{\infty} d \varepsilon_2
	  f(\varepsilon_1 )  f(\varepsilon_2 )
	  \sum_{\alpha,\alpha'} 
	  K_{\alpha,\alpha'} (\varepsilon_1,\varepsilon_2)
		 ,
  \label{eqn:e4}
\end{equation}
where
\begin{eqnarray}
 	  K_{\alpha,\alpha'} (\varepsilon_1,\varepsilon_2)
	  &=& 
	  \left<
	           \rho_{\alpha} (\varepsilon_1)
			   \rho_{\alpha'} (\varepsilon_2)
	  \right>
	  -
	  \left<
	           \rho_{\alpha} (\varepsilon_1)
	  \right>
	  \left<
			   \rho_{\alpha'} (\varepsilon_2)
	  \right> 
  \label{eqn:e5}
  \nonumber
	  \\
	 &\approx&
	  \frac{1}{2 \pi^2 V^2} \int d {\bf r}_1 \int d {\bf r}_2 
	  \mbox{Re}
	  \left[
	  \left<
	           G_{\alpha,\alpha}^{R} ({\bf r}_1,{\bf r}_1;\varepsilon_1)
	           G_{\alpha',\alpha'}^{A} ({\bf r}_2,{\bf r}_2;\varepsilon_2)
	  \right>
	 -
	  \left<
	           G_{\alpha,\alpha}^{R} ({\bf r}_1,{\bf r}_1;\varepsilon_1)
	  \right>
	  \left<
	           G_{\alpha',\alpha'}^{A} ({\bf r}_2,{\bf r}_2;\varepsilon_2)
	  \right> 
	  \right]
      .   
  \label{eqn:e6}
\end{eqnarray}
In order to calculate $K_{\alpha,\alpha'}$, we use the diagrammatic perturbation method~\cite{rf:Diagram} which is a powerful tool in the theory of disordered system.
In the language of diagrammatics, the dominant contribution to $K_{\alpha,\alpha'}$ arises from the exchange of the two Cooperon (particle-particle) ladders between two closed loops~\cite{rf:Schmid,rf:AGI} (see Fig. 1).
On the other hand, the Diffuson process gives no contribution to the average current because it has no dependence of the magnetic flux $\Phi^{em}$.
Thus we need only the Cooperon process in order to calculate the average persistent current.
As usual, we are now interested only in the field dependence resulting from interfering paths.
Therefore we ignore the less important field dependence of the  averaged single Green's function.
The two-point correlator then translates into the following expression:
\begin{eqnarray}
	  K_{\alpha,\alpha'} (\varepsilon_1,\varepsilon_2)
	  =
	  \frac{1}{2 \pi^2 V^2 \hbar^2}
	  \mbox{Re}
	  \int d {\bf x}_1
	  \int d {\bf x}_2
	  \;
	  {\cal C}_{\alpha,\alpha'} 
	  \left( {\bf x}_1,{\bf x}_2;\varepsilon_1-\varepsilon_2\right)
	  {\cal C}_{\alpha,\alpha'} 
	  \left( {\bf x}_2,{\bf x}_1;\varepsilon_1-\varepsilon_2\right)
		.
  \label{eqn:e7}
\end{eqnarray}
In this equation  we have used the definition of the particle-particle pair propagator 
\begin{eqnarray}
	  {\cal C}_{\alpha,\alpha'} 
	  \left( {\bf x}_1,{\bf x}_2; \varepsilon_1-\varepsilon_2\right)
	  =
      \frac{2 \pi \rho(0)}{\hbar}
	  \left<
	           G_{\alpha,\alpha}^{R} ({\bf x}_2,{\bf x}_1;\varepsilon_1)
	           G_{\alpha',\alpha'}^{A} ({\bf x}_2,{\bf x}_1;\varepsilon_2)
	  \right>
		,
  \label{eqn:e8}
\end{eqnarray}
where $\rho(0)$ is the density of states (per unit volume and spin) at the Fermi surface.
Following Ref. [11] we have evaluated the pair propagator $ {\cal C}_{\alpha,\alpha'}$  by using the quasi-classical Feynman path-integral method~\cite{rf:Chakraverty} and obtained
%
%
%
\begin{figure}[b]
\begin{center}
\vspace{1.0cm}
\hspace{2.5cm}
\epsfxsize=12cm
\epsfbox{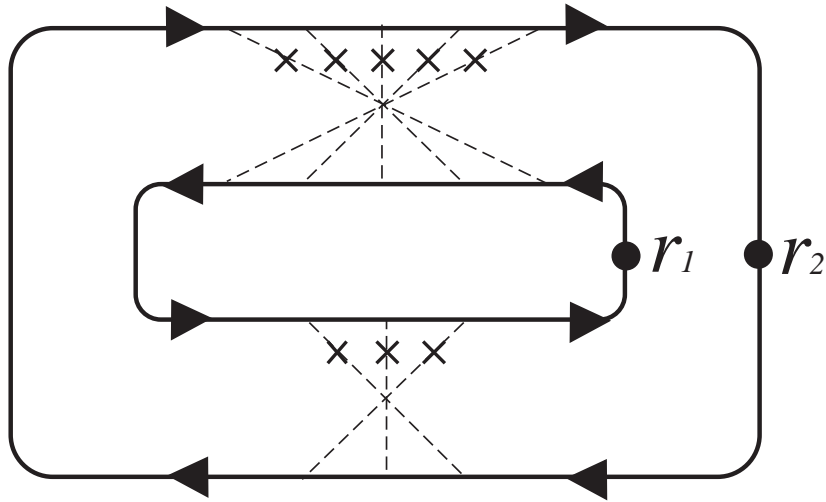}
\caption{An example of diagrams showing the leading contribution to the average persistent current.
}
\end{center}
\end{figure}
%
%
%
%
%
%
\begin{eqnarray}
	  {\cal C}_{\alpha,\alpha'} 
	  \left( {\bf x}_1,{\bf x}_2; t_1,t_2\right)
	  = 
      \theta(t_2-t_1)
	  \int_{{\bf R}(t_2)={\bf x}_2}^{{\bf R}(t_1)={\bf x}_1}
	  {\cal D} {\bf R}
	  \exp \left(
	              -i {\cal S}^{eff}_{\alpha,\alpha'} \left[ {\bf R} \right]
				 \right)
		  ,
  \label{eqn:e9}
\end{eqnarray}
where the effective action is given by 
\begin{eqnarray}
	 {\cal S}^{eff}_{\alpha,\alpha'} \left[ {\bf R} \right]
	  = 
      	  -
		   \frac{i}{4D} 
		   \int_{t_2}^{t_1} dt \left| \dot{{\bf R}} \right|^2
		  -
	       \frac{2 e}{\hbar}
		   \int_{t_2}^{t_1} dt 
		   \dot{{\bf R}}  \cdot{\bf A}^{em} \left( {\bf R}(t) \right)
		  -
		   \frac{g \mu_B B}{2 \hbar}
		   (t_2-t_1)(\alpha-\alpha')
		   -
		  \gamma_{\alpha} \left[  {\bf R} \right]
	      -
		  \gamma_{\alpha'} \left[  {\bf R} \right]
		.
  \label{eqn:e10}
\end{eqnarray}
Here the Berry phase $\gamma_{\alpha} \left[  {\bf R} \right]$ which arises from adiabatic approximation for the spin (dynamical Zeeman) propagator is of the form
\begin{eqnarray}
 \gamma_{\alpha} \left[  {\bf R} \right]
 =
 \int_{t_1}^{t_2} dt \dot{{\bf R}} \cdot {\bf A}^{g}_{\alpha} ({\bf R})
   ,
  \label{eqn:e11}
\end{eqnarray}
where ${\bf A}^{g}_{\alpha} ({\bf R})$ is the spin dependent geometric gauge potential and is given by ${\bf A}^{g}_{\alpha} ({\bf R})=(\alpha/2)\nabla \eta \left[ \cos \chi  -1 \right]$.
It should be note that the expression Eq.~(\ref{eqn:e9}) is only valid in the adiabatic regime in which the spin of the electron adiabatically follows the local direction of the non-uniform magnetic field.
This adiabaticity requires that the precession frequency $\omega_B=g\mu_B B / 2 \hbar$ is large compared to the reciprocal of the diffusion time $\tau_d=L_x^2/D$ 
($D$ is the diffusion constant) around the ring, i.e., $\omega_B \tau_d \gg 1$, or equivalently $B \gg B_c \equiv 2E_{Th}/g \mu_B$.~\cite{rf:LossDisorder,rf:adiabatic}

The path-integral representation for the pair propagator can be transformed into a differential equation,
\begin{eqnarray}
\left[
            \frac{\partial}{\partial t'} 
			+
			D
			\left\{
			         -i \frac{\partial}{\partial {\bf x}'}
					 -
		                          {\bf A}_{\alpha} \left( {\bf x}' \right)
					-
								  {\bf A}_{\alpha'} \left( {\bf x}' \right)
			\right\}^2
			-
		    i
		     \omega_B
		     (\alpha-\alpha')
\right]
 {\cal C}_{\alpha,\alpha'} 
 \left( {\bf x}_1,{\bf x}_2; t_1,t_2 \right)
 =
\delta \left( {\bf x}_1 - {\bf x}_2 \right)
\delta \left( t_1 - t_2 \right)
,
  \label{eqn:e12}
\end{eqnarray}
which  is characterized by the presence of the spin-dependent gauge potential ${\bf A}_{\alpha} \equiv (e/h){\bf A}^{em} + {\bf A}^{g}_{\alpha}$.
We now evaluated the pair propagator for the case of the metallic rings, having height $L_z$ and  thickness $L_y$.
The magnetic field was assumed to have constant magnitude, and not to vary appreciably either radially across the wall or parallel to the ring axis.
Thus the gauge potential ${\bf A}_\alpha$ can be replaced by its tangential component $A_{\alpha}^{\phi}$ ($\phi$ is the azimuthal angle). This can be expressed as $2 \pi r A_{\alpha}^{\phi} = \Phi_{\alpha} = \Phi^{em}/\Phi_0 + \alpha \Phi^g$, where $\Phi^{em}$ is the electromagnetic flux through the area $\pi r^2$ and the geometric flux is given by
\begin{eqnarray}
   \Phi^g
   =
   \frac{1}{4 \pi} 
   \int_0^{2 \pi}
   d \phi
   \left[ 
    \cos \chi (\phi)
	-
	1
   \right]
   \partial_{\phi} \eta(\phi)
   .
  \label{eqn:e13}
\end{eqnarray}
The geometric flux take a simple form in the case of a cylindrically symmetric texture ($\eta=\phi$), i.e.,  $\Phi^g=(\cos \chi -1)/2$, where $\chi$ is the constant tilt angle away from the cylinder axis.
According to the standard method,~\cite{rf:AronovSharvin} we can show that the pair propagator ${\cal C}_{\alpha,\alpha'}$ is explicitly given by~\cite{rf:LossDisorder} 
\begin{eqnarray}
	 {\cal C}_{\alpha,\alpha'}
	 \left( {\bf x}_1,{\bf x}_2;t_1,t_2\right)
	  &=& 
	  \frac{1}{V}
	  \sum_{{\bf k}} 
	  \int_{-\infty}^{\infty} 
	  \frac{d \omega}{2 \pi}
	  {\cal C}_{\alpha,\alpha'}
	  \left( {\bf k} , \omega \right)
	  e^{
	        i {\bf k} {\bf \cdot} ({\bf x}_1-{\bf x}_2)
			-
			 i \omega (t_1-t_2)
		   }
	  , 
  \label{eqn:e14}
	  \\
	  {\cal C}_{\alpha,\alpha'}
	  \left( {\bf k} , \omega \right)
	 &=&
	  \frac{1}
	            {
				  \displaystyle{
				  -i ( \omega - \varpi_{\alpha,\alpha'})
				  +
				  \frac{1}{\tau_{\varphi}}
				  +
				  D {\bf k}_\perp^2
				  +
				  \frac{D}{r^{2}}
				  \left[
				             n_x - (\Phi_{\alpha} + \Phi_{\alpha'})
				  \right]^2
				  }
				}
	 ,
  \label{eqn:e15}
\end{eqnarray}
where ${\bf k}=(2 \pi n_x /L_x, {\bf k}_\perp)$, ${\bf k}_\perp=(2 \pi n_y / L_y,2 \pi n_z/L_z)$ ($n_x,n_y,n_z$ are integer) and $ \varpi_{\alpha,\alpha'} \equiv \omega_B (\alpha-\alpha')$.
For convenience we have incorporated the effect of dephasing $1/\tau_\varphi=D/L_{\varphi}^2$
($L_{\varphi}$ is the phase coherence length) directly into the pair propagator.
Note that for diagonal component of the pair propagator ${\cal C}_{\alpha,\alpha}$, the Zeeman terms cancel, i.e., $\varpi_{\alpha,\alpha}=0$.

In the following we shall calculate the diagonal part ($\alpha'=\alpha$) and the off-diagonal part ($\alpha'=-\alpha$) of the average current separately, namely,
\begin{eqnarray}
     \left< I (\Phi^{em}) \right> 
	 \equiv 
	 \left< I^D (\Phi^{em}) \right> +\left< I^{OD} (\Phi^{em}) \right>
     .
  \label{eqn:e16}
\end{eqnarray}
The diagonal part is attributed to the interference between same spin states.
If the thickness and the height of the ring is small compared with the length $L_\varphi$, then the integration over ${\bf k}_\perp$ should be replaced by a summation with only the term corresponding to ${\bf k}_\perp= {\bf 0}$ retained.
Substituting the Fourier transform of Eq.~(\ref{eqn:e15}) into Eq.~(\ref{eqn:e7}) and using the Matsubara method, therefore, we obtain the diagonal part of the average current 
\begin{eqnarray}
 	 \left< I^D (\Phi^{em}) \right>
	 =
	 - \frac{\Delta}{2 \pi \beta}
	 \sum_{\alpha=\pm 1} \mbox{Re} \frac{\partial}{\partial \Phi^{em}}
	 \sum_{\nu_\ell>0}  
	 \sum_{n_x=-\infty} ^{\infty}
	 \frac{\nu_\ell}{
	                \left\{
					\displaystyle{
	                \nu_\ell
					+
					\frac{\hbar}{\tau_\varphi}
					+
					4 \pi^2 E_{Th}
					\left( n_x + 2 \Phi_\alpha \right)^2
					}
					\right\}^2
					}	
					,
  \label{eqn:e17}
\end{eqnarray}
where $\beta=1/k_BT$ and $E_{Th}=\hbar D/L_x^2$ is the Thouless energy and $\nu_\ell=2\pi \ell /\beta$ ($\ell$ is an integer) is the boson Matsubara frequency, respectively.
In the limit of zero temperature, the Matsubara sum turns into an integral $\sum\nolimits_{\nu} \to 2 \pi/\beta  \int d \nu$, which is easily evaluated.
This yields for the diagonal part of the averaged persistent current at $T=0$ as
\begin{eqnarray}
 	 \left< I^D (\Phi^{em}) \right>
	 =
	 \frac{I_0}{M}
	 \sum_{\alpha=\pm 1} 
	 \sum_{n=1}^{\infty} 
	 \exp \left( 
	                      -n \frac{L_x}{L^B_{\varphi}} 
	           \right)
	 \sin \left( 4 \pi n \Phi_\alpha \right)
	        .
  \label{eqn:e18}
\end{eqnarray}
$I_0=e \upsilon_F/L_x$ is the current carried by a single electron state in an ideal one-dimensional ring and  $M=k_F^2V/L_x$ is the effective channel number, where $\upsilon_F$ is the Fermi velocity and $k_F=m\upsilon_F/\hbar$ is the Fermi wave number.
In Eq.~(\ref{eqn:e18}) the magnetic dephasing length $L^B_{\varphi}$ is given by 
$
1/(L^B_{\varphi})^2 = 1/(L_{\varphi})^2+1/(L^B)^2
$
and $L^B= \sqrt{3} \Phi_0 / (2 \pi L_y \left| B_z \right|)$ accounts for the dephasing effect of the magnetic field penetrating into the sample region,~\cite{rf:AltshulerAronov} where $B_z$ is the $z$ component of ${\bf B}$.
Therefore we will observe the persistent current oscillation with somewhat reduced amplitude.~\cite{rf:BerryWL}
It is important to note that unlike the case of the one dimensional ring~\cite{rf:LossPC} the inhomogeneous magnetic fields leads to the dephasing effect to the persistent current.

For the off-diagonal part of the persistent current which results from the interference between opposite spin states, same procedure as above yields
\begin{eqnarray}
	 \left< I^{OD} (\Phi^{em}) \right>
	 &=&
	 - \frac{\Delta}{\pi \beta}
	 \mbox{Re} \frac{\partial}{\partial \Phi^{em}}
	 \sum_{\nu_\ell>0}  
	 \sum_{n_x=-\infty} ^{\infty}
	 \frac{\nu_\ell}
	           {
	 			    \displaystyle{
	                \left\{
	                       \nu_\ell
					      +
					      \frac{\hbar}{\tau_\varphi}
					      +
					      4 \pi^2 E_{Th}
					      \left( n_x + 2 \frac{\Phi^{em}}{\Phi_0} \right)^2
					  \right\}^2
					  +
					  \left(
					            g \mu_B B 
					   \right)^2
					}
					}
  \label{eqn:e19}
	 \\
	 &=&
	 \frac{I_0}{M}
	 \sum_{n=1}^{\infty} 
	 i_n
	 \sin \left( 4 \pi n \frac{\Phi^{em}}{\Phi_0} \right)
	        ,
  \label{eqn:e20}
\end{eqnarray}
with 
\begin{eqnarray}
	 i_n
	 &=&
	 2 n \xi
	 \int_0^{
	                  \sqrt{  \sqrt{\Gamma^2 +1} -\Gamma }
	                }
	 d y
	 \left(
	              \frac{1-y^4}{2y^2} - \Gamma
	 \right)
	 \left[
	            \cos(n \xi y) +\frac{1}{y^2} \sin(n \xi y)
	 \right]
	 \exp \left( -n \frac{\xi}{y} \right)
	        ,
  \label{eqn:e21}
\end{eqnarray}
where $\Gamma \equiv D/[2 \omega_B (L_\varphi^B)^2]$ and $\xi \equiv \sqrt{\hbar \omega_B/E_{Th}}$.
Note that the off-diagonal term does not depend on the Berry phase.
In contrast to the diagonal term, we can expect that this term gives negligibly small contributions to the average current in the adiabatic regime because the Cooperon pole has a non-zero Zeeman term.  
Figure 2 shows the $B$ dependence of $\left< I^{D} \right>$ and  $\left< I^{OD} \right>$ of quasi-one-dimensional rings ($L_x=11\mu$m and  $L_y=2.0\mu$m) in the case of the cylindrically symmetric texture.
As expected above, $\left< I^{OD} \right>$ is much smaller than $\left< I^{D} \right>$ in the adiabatic regime, $B \gg B_c \approx 10$Gauss.
Therefore the average persistent current is given by
\begin{eqnarray}
 	 \left< I (\Phi^{em}) \right>
	 \approx
	 \left< I^D (\Phi^{em}) \right>
     =
	 \frac{I_0}{2M}
	 \sum_{\alpha=\pm 1} 
	 \frac{
	             \displaystyle{
	             \sin \left(  4 \pi \Phi_\alpha \right)
				 }
	           }
			   {
			    \displaystyle{
				\cosh \left( \frac{L_x}{L^B_{\varphi}} \right)
				-
				\cos \left(  4 \pi \Phi_\alpha \right)
				}
				}
  \label{eqn:e22}
\end{eqnarray}
which is a periodic function of $\Phi^{em}$ and $\Phi^g$ with period $\Phi_0/2$ and $1/2$, respectively.
These periods are half the value of the case of a single one-dimensional ring.~\cite{rf:LossPC}
As the half flux periodicity of the electromagnetic flux in disordered rings,~\cite{rf:AGI} that of geometric flux is ascribed to the ensemble averaging for fixed particle number: averaging eliminates the first Fourier component of the current although the second components which results from the interference between time-reversed trajectories survives.
If the magnetic field is homogeneous then the Berry phase vanishes, i.e., $\Phi^g=0$.
In this case the expression Eq.~(\ref{eqn:e22}) is reduced to the well-known result.~\cite{rf:Schmid,rf:AGI}

In summary, we have investigated the persistent current of disordered rings in the inhomogeneous magnetic field.
Halving of the geometric flux period with respect to the single ballistic one-dimensional ring~\cite{rf:LossPC} is obtained in the canonical disorder-ensemble average.
We have also showed that the existence of magnetic field which penetrates the sample causes the dephasing effect.

Finally we point out that the phenomena investigated in this paper will also appear in doubly-connected chaotic billiards.~\cite{rf:chaos}
We expect that these systems will open a new area to explore the effect of $chaos$ on the Berry phase.

The author wishes to thank B. A. Friedman for helpful comments.\\

%
%
.
%
%
%
%
%
%
%
\begin{figure}[t]

\vspace{-3.0cm}
\hspace{3.3cm}
\epsfxsize=10.0cm
\epsfbox{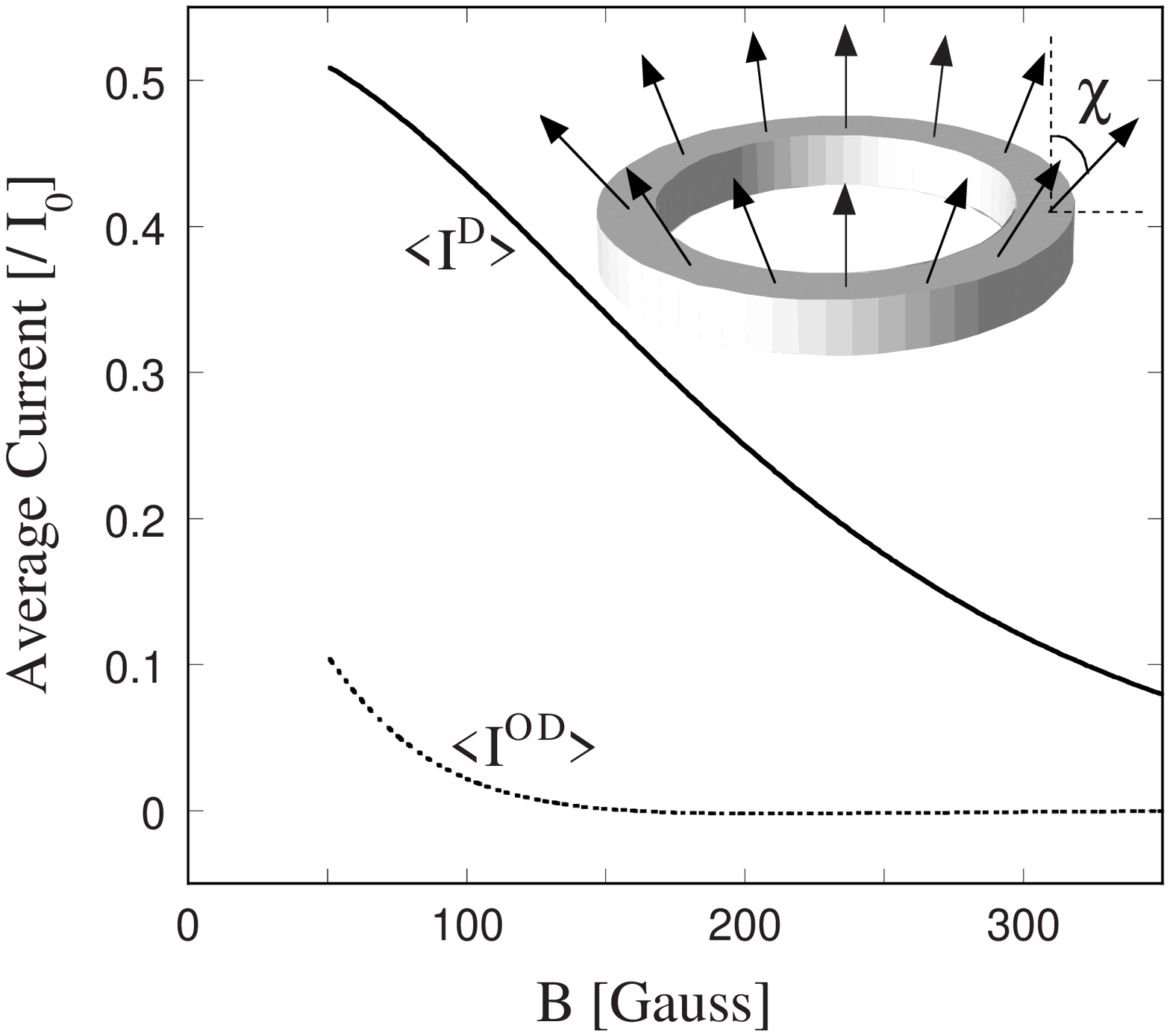}

\caption{Magnetic field dependence of the persistent current of quasi-one-dimensional disordered rings embedded in the cylindrically symmetric texture with a fixed value of $\Phi^{em}/\Phi_0=1/8$ and $\chi=\pi/3$.
Plotted is the result for $D=9.0  \times10^{-3} m^2$/s and $L_\varphi=5.5\mu m$. [17]
The solid (dotted) line represents the diagonal (off-diagonal) part of the persistent current. 
Inset shows a schematic drawing of a disordered ring in inhomogeneous magnetic fields (arrows).
}
\end{figure}
%
%
%
%
%
%
%
%
%
%

%
%
%
%
%
%
%
%
%
%
%
%
%
%
%
%
%
%
%
%
\end{document}